\documentstyle[12pt,twoside]{article}

\catcode `\@=11
\@addtoreset{equation}{section}

\def\section{\@startsection {section}{1}{\z@}{3ex plus 1ex minus
.2ex}{1.2ex plus .2ex}{\normalsize\bf}}

\def\subsection{\@startsection{subsection}{2}{\z@}{3ex plus 1ex minus
.2ex}{1.2ex plus .2ex}{\normalsize\bf}}

\def\thebibliography#1{\section*{{
                                    \rm REFERENCES}\@mkboth
 {BIBLIOGRAPHY}{BIBLIOGRAPHY}}\list
 {[\arabic{enumi}]}{\settowidth\labelwidth{[#1]}\leftmargin\labelwidth
 \advance\leftmargin\labelsep\usecounter{enumi}}
 \def\newblock{\hskip .11em plus .33em minus -.07em}
 \sloppy\clubpenalty4000\widowpenalty4000
 \sfcode`\.=1000\relax}

\catcode `\@=12

\title{\vspace*{1.5cm} \normalsize\bf 
ON FIELD THEORETIC GENERALIZATIONS OF A POISSON 
ALGEBRA
}

\author{\vspace{1.5cm} \sc Igor V. Kanatchikov\thanks{e-mail: 
kai@fuw.edu.pl} \vspace{-1.5cm}\\    
\small Laboratory of Analytical Mechanics and Field Theory \vspace{-0.2cm}\\ 
\small Institute of Fundamental Technological Research\vspace{-0.2cm} \\
\small Polish Academy of Sciences\vspace{-0.2cm}\\
\small \'Swi\c etokrzyska 21,  Warszawa  PL-00-049, Poland 
}

\date{\small \it (Submitted April 1997   ------  Accepted July 1997) }

\begin{document}

\maketitle 

\markboth{\centerline{\small  I.V. KANATCHIKOV}}{ \hspace*{-12.5pt}
\centerline {\small ON FIELD THEORETIC   
GENERALIZATIONS OF A POISSON ALGEBRA}}


\newcommand{\beq}{\begin{equation}}
\newcommand{\eeq}{\end{equation}}
\newcommand{\beqa}{\begin{eqnarray}}
\newcommand{\eeqa}{\end{eqnarray}}
\newcommand{\nn}{\nonumber}

\newcommand{\half}{\frac{1}{2}}

\newcommand{\xt}{\tilde{X}}

\newcommand{\uind}[2]{^{#1_1 \, ... \, #1_{#2}} }
\newcommand{\lind}[2]{_{#1_1 \, ... \, #1_{#2}} }
\newcommand{\com}[2]{[#1,#2]_{-}} 
\newcommand{\acom}[2]{[#1,#2]_{+}} 
\newcommand{\compm}[2]{[#1,#2]_{\pm}}

\newcommand{\lie}[1]{\pounds_{#1}}
\newcommand{\co}{\circ}
\newcommand{\sgn}[1]{(-1)^{#1}}
\newcommand{\lbr}[2]{ [ \hspace*{-1.5pt} [ #1 , #2 ] \hspace*{-1.5pt} ] }
\newcommand{\lbrpm}[2]{ [ \hspace*{-1.5pt} [ #1 , #2 ] \hspace*{-1.5pt}
 ]_{\pm} }
\newcommand{\lbrp}[2]{ [ \hspace*{-1.5pt} [ #1 , #2 ] \hspace*{-1.5pt} ]_+ }
\newcommand{\lbrm}[2]{ [ \hspace*{-1.5pt} [ #1 , #2 ] \hspace*{-1.5pt} ]_- }
\newcommand{\pbr}[2]{ \{ \hspace*{-2.2pt} [ #1 , #2 ] \hspace*{-2.55pt} \} }
\newcommand{\we}{\wedge}
\newcommand{\dv}{d^V}
\newcommand{\nbrpq}[2]{\nbr{\xxi{#1}{1}}{\xxi{#2}{2}}}
\newcommand{\lieni}[2]{$\pounds$${}_{\stackrel{#1}{X}_{#2}}$  }

\newcommand{\rbox}[2]{\raisebox{#1}{#2}}
\newcommand{\xx}[1]{\raisebox{1pt}{$\stackrel{#1}{X}$}}
\newcommand{\xxi}[2]{\raisebox{1pt}{$\stackrel{#1}{X}$$_{#2}$}}
\newcommand{\ff}[1]{\raisebox{1pt}{$\stackrel{#1}{F}$}}
\newcommand{\dd}[1]{\raisebox{1pt}{$\stackrel{#1}{D}$}}
\newcommand{\nbr}[2]{{\bf[}#1 , #2{\bf ]}}
\newcommand{\der}{\partial}
\newcommand{\oo}{$\Omega$}
\newcommand{\Om}{\Omega}
\newcommand{\om}{\omega}
\newcommand{\eps}{\epsilon}
\newcommand{\si}{\sigma}
\newcommand{\Lm}{\bigwedge^*}

\newcommand{\inn}{\hspace*{2pt}\raisebox{-1pt}{\rule{6pt}{.3pt}\hspace*
{0pt}\rule{.3pt}{8pt}\hspace*{3pt}}}
\newcommand{\sro}{Schr\"{o}dinger\ }
\newcommand{\bm}{\boldmath}
\newcommand{\vol}{\omega}

\newcommand{\bd}{\mbox{\bm $d$}}
\newcommand{\bder}{\mbox{\bm $\der$}}
\newcommand{\bI}{\mbox{\bm $I$}}


\newcommand{\oldtitle}{
\begin{document}\vspace*{2.5cm}
\noindent
\begin{center}
{\bf ON FIELD THEORETIC GENERALIZATIONS OF A POISSON 
ALGEBRA
}\vspace{1.3cm}\\
\end{center}
\noindent
\hspace*{1in}
\begin{minipage}{13cm}
\makebox[3mm]{ }Igor V. Kanatchikov
\vspace{0.3cm}\\
\makebox[3mm]{ }Laboratory of Analytical Mechanics and Field Theory \\  
\makebox[3mm]{ }Institute of Fundamental Technological Research \\
\makebox[3mm]{ }Polish Academy of Sciences \\
\makebox[3mm]{ }Swi\c etokrzyska 21, Warszawa PL-00-049 \\
      \makebox[3mm]{ }Poland \\
\end{minipage}

\vspace*{0.5cm}
}
\vspace*{-96mm}
\hbox to 6.2truein{
\footnotesize\it 
to appear in   
\hfil \hbox to 0 truecm{\hss 
\normalsize\rm July 1997 }\vspace*{-1mm}}
\hbox to 6.2truein{
\vspace*{-1mm}\footnotesize 
Rep. Math. Phys. 
\hfil 
} 
\hbox to 6.2truein{
\vspace*{-1mm}\footnotesize 
vol. {\bf 40} (1997) p.225  \hfil 
\hbox to 0 truecm{ 
\hss \normalsize hep-th/9710069} 
}
\vspace*{72mm}


\begin{flushright} 
\begin{minipage}{5.3in}
{\footnotesize 
A few generalizations of a Poisson algebra to field theory 
canonically formulated in terms of the 
polymomentum variables are discussed.  
A graded Poisson bracket on differential forms and 
an $(n+1)$-ary bracket on functions are considered. 
The Poisson bracket on differential forms 
gives rise to various generalizations of a Gerstenhaber algebra: 
the noncommutative (in the sense of Loday) and the higher-order 
(in the sense of the higher order graded Leibniz rule). 
The $(n+1)$-ary bracket fulfills the properties of the Nambu bracket 
including the ``fundamental identity'', thus leading 
to the Nambu-Poisson algebra. We point out that in the field theory context 
the Nambu bracket with a 
covariant analogue of 
Hamilton's function determines a {\em joint} evolution of several 
dynamical variables.  
}   
\end{minipage}
\end{flushright} 

\section{Introduction }

The purpose of the present paper is to discuss  
generalizations 
of a Poisson algebra in field theory which appear 
within the so-called De Donder--Weyl (DW) canonical theory 
for fields, the essence of which is briefly recalled below 
(see also  \cite{ikanat0} and  references therein). 
The DW theory is in a sense a manifestly  space-time 
symmetric extension of the Hamiltonian formulation 
to field theory, which  belongs  
to the Lepagean class of canonical theories 
known from the calculus of variations 
(see \cite{lepage}). All these formulations are 
referred  to as {\em polymomentum} here, 
as  their common feature is that variables, called  polymomenta, 
similar to conjugate momenta, are associated to every space-time 
derivative of fields. 

Our interest to the  above generalizations 
is motivated by the question 
as to whether the DW formulation, 
or another polymomentum formulation,  
can  provide us with a starting point 
for a  certain 
``canonical quantization'' 
scheme in field theory. 
Such a scheme would certainly 
be of interest because 
in accordance with 
the general features of 
polymomentum formulations 
it would not assume  
an explicit distinction between 
the space and time variables,  
as the standard Hamiltonian formalism does,    
and potentially may not suffer from the 
functional analytic difficulties related 
to the infinite dimensionality of the latter. 
A study of analogues of a Poisson algebra in the context 
of 
polymomentum formulations 
seems to be a natural 
preliminary step towards such a quantization. 

The present discussion can be viewed as a continuation of our 
earlier reports (see \cite{ikanat0,ikanat1,ikanat2}). 
Here we partially review  our previous results 
and partially present some new ones. 
The latter include  the Gerstenhaber algebra structure with 
respect to the ``co-exterior product'' in Sect. 2 
and a discussion 
of the Nambu bracket in field theory in Sect. 4.

Let us start by recalling what  
the DW polymomentum formulation is. 
In  field theory 
given by the Lagrangian density 
$L=L(y^a, \der_i y^a, x^j)$, where 
$\{y^a \}$ are 
field variables, 
$\{ \der_i y^a \}$ are 
their space-time derivatives 
and $\{x^i \}$ are 
space-time coordinates,  
 the set of variables 
$p_a^i:=\der L/ \der (\der_i y^a)$ 
(= {\em polymomenta}) 
and the quantity  
$H:= \der_i y^a p_a^i - L$ 
(= the {\em DW Hamiltonian function}) 
can be introduced 
if det$||\der^2 L/\der (\der_i y^a) \der (\der_j y^b)||$$\neq 0$.  
This  allows us to write the Euler--Lagrange field equations 
in 
the manifestly covariant first order form 
\beq
\der_i y^a = \der H / \der p^i_a, \quad \der_i p^i_a = - \der H/ \der y^a 
\eeq
which can naturally be viewed as a multidimensional 
(or rather ``multi-time'') 
generalization of Hamilton's canonical equations. 
They  are  referred to as 
the  
 {\em DW Hamiltonian equations. }  
The arena of field  dynamics is now the finite dimensional 
{\em polymomentum phase space} 
of variables $z^M:=(y^a, p_a^i, x^i)$ and all the space-time variables 
enter on an equal footing as analogues of the time variable in mechanics. 
We have argued earlier \cite{ikanat0,ikanat1} that 
the proper analogue of Poisson brackets  can be obtained from the 
following generalization of the symplectic 
form to the polymomentum phase space 
which we call  the {\em polysymplectic form}:  
\beq
\Omega := dy^a\we dp^i_a \we \omega_i, 
\eeq
where $\omega :=dx^1\wedge ... \wedge dx^n$ 
is the volume-form on the space-time manifold, 
$\omega_i:=\der_i \inn \omega $, 
and $\inn\ $ denotes the inner product of a 
(multi)vector with a form. 
A possible intrinsic meaning of $\Omega $ as 
a representative of a certain equivalence class 
is discussed in \cite{ikanat2}.  
In the following the variables $z^v=(y^a, p_a^i)$ and $x^i$, 
as well as the corresponding subspaces, 
are referred to as 
vertical and horizontal respectively. We also use the notation 
$\der\lind{M}{p}:=\der_{M_1}\we ... \we \der_{M_p}$. 

\section{Hamiltonian forms}

The usual definition of the Poisson bracket using the symplectic form  
can be generalized to the DW polymomentum formulation in field theory 
as follows (see \cite{ikanat0,ikanat1} for more details). 
Let us consider the map of horizontal forms of degree $p$, $p=0,...,n-1$ 
$$\ff{p}:=\frac{1}{(n-p)!}F\uind{i}{n-p}(z)\der\lind{i}{n-p}
\inn\omega,
$$  
to vertical multivectors of degree $(n-p)$,
$$\xx{n-p}=\frac{1}{(n-p)!} 
[ X^a{}\uind{i}{n-p-1}\der_a{}\lind{i}{n-p-1} + 
X_a^i{}\uind{i}{n-p-1}   \der^a_i{}\lind{i}{n-p-1} + ... ] 
$$ 
which is given by 
\beq
\xx{n-p}{}_F\inn\ \Omega= \dv \ff{p} .
\eeq
Here a multivector of degree $q$ is called vertical if it 
annihilates any horizontal $q$-form, 
whence the component expression above follows; 
the higher vertical components of the multivector 
are omited as they play no role here.   
The vertical exterior 
differential, $\dv$,  
of a $p$-form  is given in components by 
$$d^V \ff{p} = \frac{1}{(n-p)!}\der_v F\uind{i}{n-p}dz^v\we 
\der\lind{i}{n-p}\inn \omega.$$   
The  {}   horizontal forms (and vertical multivectors) 
for which the map  above exists are called {\em Hamiltonian. }  
The  map (2.1) is equivalent to the following  
relations between the 
components of 
a $p$-form $\ff{p}$ 
and those of the associated $(n-p)$-multivector $\xx{n-p}{}_F$:  
\beqa
(n-p) 
X_a^i{}\uind{i}{n-p-1}&=&  \der_a F\uind{i}{n-p-1}{}^i ,        \\
-(n-p) X^{a[}{}\uind{i}{n-p-1}\delta^{i]}_j
&=&
\der_j^a F\uind{i}{n-p-1}{}^i . 
\eeqa  
The latter relation obviously imposes a restriction 
on forms fulfilling 
eq. (2.1). Namely, 
an analysis of its consistency conditions shows 
that Hamiltonian forms are restricted to 
the 
specific polynomials of polymomenta with 
coefficients 
depending on the space-time and field variables, i.e. (for $p\neq 0,n$) 
\beq
\ff{p}=\frac{1}{(n-p)!}
\sum^{n-p}_{k=0}p^{i_1}_{a_1} ... p^{i_k}_{a_k} 
f\uind{a}{k}{}^{i_{k+1} ... i_{n-p}} (x^i,y^a) \der\lind{i}{n-p}\inn \omega .
\eeq 
The Poisson bracket of two Hamiltonian forms is now defined as follows: 
\beq
\pbr{\ff{p}{}_1}{\ff{q}{}_2}:= (-1)^{n-p}X_{F_1}\inn X_{F_2} \inn \Omega .
\eeq
It is easy to show that the bracket in (2.5) fulfills the axioms of 
a graded Lie algebra: 
\mbox{graded anticommutativity } 
\beq
\pbr{\ff{p}_1}{\ff{q}_2} = -(-1)^{g_1 g_2}
\pbr{\ff{q}_2}{\ff{p}_1}, 
\eeq
\mbox{and the graded Jacobi identity }  
\nn \\
\beq
\mbox{$(-1)^{g_1 g_3} \pbr{\ff{p}}{\pbr{\ff{q}}{\ff{r}}}$} 
+
\mbox{$(-1)^{g_1 g_2} \pbr{\ff{q}}{\pbr{\ff{r}}{\ff{p}}}$} 
+
\mbox{$(-1)^{g_2 g_3} \pbr{\ff{r}}{\pbr{\ff{p}}{\ff{q}}}= 0,$}  
\eeq
where $g_1 = n-p-1$,  $g_2 = n-q-1$,  $g_3 = n-r-1$ are 
degrees of  
the 
corresponding forms with respect to the bracket operation. 

Further, from eq. (2.4) it is clear that the space of Hamiltonian forms is 
not 
 closed 
with respect to the exterior product. 
However, another graded commutative associative product of horizontal 
forms can   be constructed with respect to which the space of Hamiltonian 
forms {\em is} 
closed.  
This product operation $\bullet$ 
(let us call it the {\em co-exterior product}) 
is given by the formula 
\beq
F\bullet G:=*^{-1}(*F\we *G).
\eeq 
Clearly, ${\rm deg}(\ff{p}\bullet \ff{q})= p+q-n$ 
and $\ff{p}\bullet \ff{q}=(-1)^{(n-p)(n-q)}\ff{q}\bullet \ff{p}$. 
Note that 
only the volume $n$-form $\omega$ is needed in order to define the 
co-exterior product, not the metric structure. 
Now, a remarkable fact is that the Poisson bracket defined in (2.5) 
fulfills the graded Leibniz rule with respect to the co-exterior product
\beq
\pbr{\ff{p}}{\ff{q}\bullet \ff{r}}=
\pbr{\ff{p}}{\ff{q}}\bullet \ff{r} 
+ (-1)^{(n-q)(n-p-1)}\ff{q}\bullet\pbr{\ff{p}}{\ff{r}}. 
\eeq
The proof is based on a straightforward componentwise calculation. 

	Eqs. (2.6), (2.7) and (2.9) lead us to the conclusion 
that the space of Hamiltonian forms is a Gerstenhaber algebra  (see e.g. 
\cite{gerst}) with 
respect to the graded bracket operation $\pbr{}{}$ and the 
co-exterior product $\bullet$. 
An extension of the bracket and the corresponding 
algebraic structure to arbitrary horizontal forms 
is discussed in 
	\mbox{Sect. 3.} 

Note that the bracket above  allows us to   
write  the equations of motion in  
Poisson bracket formulation (see [1]).   
For example, the DW Hamiltonian  equations (1.1)  
take the following form:   
\beq
*^{-1}{\bd}(y^a\omega_i)=\pbr{H}{y^a\omega_i}, 
\quad 
*^{-1}{\bd} (p_a^i\omega_i)=\pbr{H}{p_a^i\omega_i}, 
\eeq 
where ${\bd}$ is the total exterior differential, such that 
${\bd (F^i\omega_i):=\der_v F^i \frac{\der z^v}{\der x^j} dx^j\we \omega_i}$. 
Note also that  $(n-1)$-forms $p_a^i\omega_i$ are canonically 
conjugate  to 
field variables $y^a$ in the sense that 
$\pbr{p_a^i\omega_i}{y^b}=\delta^b_a$.

\section{Non-Hamiltonian forms and a noncommutative 
Gerstenhaber algebra}

Horizontal forms which cannot be mapped to vertical 
multivectors are called non-Hamiltonian. 
These forms, however, can be mapped to more general graded differential 
operators acting on the exterior algebra of forms.  
Multivectors are 
just a particular case of the latter.  
In general,  graded differential operators are 
represented by multivector valued forms. 
In fact, the map 
\beq
\xt_F \inn \Omega = \dv \ff{p} 
\eeq
always exists if $\xt_F$ is taken to be a vertical multivector valued 
(horizontal) one-form: 
$$\xt = X^v{}\uind{i}{n-p}{}_k dx^k\otimes \der_v{}\lind{i}{n-p}.$$
The ``interior product'' 
$\inn$ in (3.1)  means the substitution of $\xt$ into the form 
$\Omega$, 
that is the inner product with the multivector part of $\xt$ 
is supposed to act first and then the exterior product with the 
covector part follows. 
Obviously the degree of the 
(operation of the substitution of the) 
operator $\xt$ above is $-(n-p)$.  

As before, the bracket of two forms can be defined as follows
\beq
\pbr{\ff{p}_1}{\ff{q}_2}:=(-1)^{n-p} \xt_1 \inn \dv \ff{q}_2 . 
\eeq
However, now the bracket above will lack the graded anticommutativity  
for  the graded commutator of operators $\xt_1$ and $\xt_2$  
does not vanish unless both are representable  by multivectors. 
Nevertheless, we still arrive here at 
a very interesting algebraic structure   
which is related to a noncommutative generalization of Lie algebra 
introduced \nolinebreak by Loday \cite{loday} under the name of 
Leibniz algebra  
(we  use the name  ``Loday algebra'' instead).  

Note first that all operators $\xt$ for which eq. (3.1) is fulfilled 
obey  the relation  
$L_{\xt}\Omega=0$ (i.e. they are ``locally Hamiltonian'' in a sense), 
where 
$L_{\xt}:=[\xt,\dv]:=\xt \co \dv - (-1)^{|\xt|}\dv \co \xt$ 
is a generalized Lie derivative; $|\xt|$ denotes the degree of the 
graded operator $\xt$ and $[\;,\;]$ is a graded commutator. 
For any two graded operators  
a differential bracket operation 
(an analogue of the Lie bracket of vector 
fields)  can be defined 
 \beq
\lbr{\xt_1}{\xt_2}:=[L_{\xt_1},{\xt_2}]=L_{\xt_1}\co  {\xt_2}- 
(-1)^{|{\xt_2}|(|\xt_1|+1)} {\xt_2}\co L_{\xt_1}. 
\eeq
Then it becomes obvious that the bracket in (3.2) is in fact induced 
by the differential bracket of graded operators associated with forms 
according to (3.1), i.e. 
\beq
 \lbr{\xt_F}{\xt_G}\inn \Omega=-\dv \pbr{F}{G} .
\eeq
Now,  the following identities for the bracket (3.2) can be 
proved: 

(i) the left graded Loday identity: 
\beq
\pbr{\pbr{F}{G}}{K}=
\pbr{F}{\pbr{G}{K}}- (-1)^{(n-F-1)(n-G-1)}
\pbr{G}{\pbr{F}{K}} 
\eeq
and (ii) the right graded  Leibniz rule:
\beq
\pbr{F\we G}{K}
=F\we \pbr{G}{K} + (-1)^{G(n-K-1)}\pbr{F}{K}\we G . 
\eeq

{\em Proof:}
\noindent 
Here we  omit tildes over  graded operators. 
Capital letters in the 
exponents of the minus signs denote 
the exterior degree of the corresponding forms.    
We also use some easy to reveal 
properties of  the operations 
introduced above. 

First let us prove (3.5):  
\beqa
&&\pbr{\pbr{F}{G}}{K}=(-1)^{n-(F+G-n+1)}L_{X_{\pbr{F}{G}}} K
= - (-1)^{n-(F+G-n+1)}L_{\lbr{X_F}{X_G}} K
\nn \\
&& =(-1)^{2n-F-G}\lbr{X_F}{X_G}\inn \dv K
=(-1)^{2n-F-G}[L_{X_F},X_G]\dv K
\nn \\
&&
=(-1)^{2n-F-G}\left (L_{X_F}X_G \inn \dv K + (-1)^{(n-G)(n-F-1)}X_G \inn L_{X_F}
\right )
\dv K
\nn 
\\
&&
=\pbr{F}{\pbr{G}{K}}
-(-1)^{(n-F-1)(n-G-1)}\pbr{G}{\pbr{F}{K}} .
\eeqa
In order to prove  (3.6) 
let us  construct the operator 
associated with the exterior    
product of two forms $F$ and $G$. It obeys 
\beqa 
X_{F\we G}\inn \Omega 
&=& \dv (F\we G)=
(-1)^{G(F+1)}G\we \dv F + (-1)^F F \we \dv G
\nn \\
&=&[(-1)^{G(F+1)}G \co X_F + (-1)^F F \co X_G]\inn \Omega,   
\nn 
\eeqa 
so that 
\beq
X_{F\we G} =  (-1)^F F \co X_G + (-1)^{G(F+1)}G \co X_F 
\eeq
whence (3.6) follows. {\em q.e.d.}

\medskip 

It is interesting to note that the graded anticommutativity of the 
bracket is replaced now by a weaker condition which is a consequence of 
(3.5) 
\beq
\pbr{\pbr{F}{G}}{K}=- (-1)^{(|F|+1)(|G|+1)} \pbr{\pbr{G}{F}}{K}.  
\eeq

The structure which appeared  here generalizes the known structure of 
a Gersten\-haber algebra [4]. 
Namely, the axioms of graded anticommutativity 
and graded Jacobi identity are weakened  to the (left) graded Loday 
identity, eq. (3.5), 
and the graded derivation property is valid only in the sense of 
the {\em right}  Leibniz rule.  The corresponding structure can  naturally 
be called 
a 
{\em noncommutative (right) Gerstenhaber algebra}. 

Let us consider now an analogue of the left Leibniz rule with respect to 
the exterior product. One cannot expect the graded Leibniz rule to be 
fulfilled here because 
multivector valued forms are not graded derivations but rather higher-order 
graded differential operators on the exterior algebra which are composed from 
derivations given, according to the Fr\"olicher-Nijenhuis theorem 
\cite{fn}, by vectors and vector valued forms. 
Consequently, an analogue of the Leibniz rule for  higher-order differential 
operators will appear here. 
This property is similar to the ``second order Leibniz rule'' for the 
operator of second derivative 
\beq
(abc)''=(ab)''c+(ac)''b+(bc)''a-a''bc-ab''c-abc''.
\eeq
To formulate the higher-order graded analogue of the property above 
let us recall Koszul's characterization of  higher-order graded 
differential operators on (graded) commutative algebras \cite{koszul}. 
Given an operator $D$ on the algebra $\bigwedge^*$ 
one can construct a set of $r$-linear maps associated with it,  
$\Phi{}^r_D: \bigotimes{}^r \Lm \rightarrow \Lm$, given by 
\beq
\Phi{}^r_D(F_1, ..., F_r) := 
m \co (D\otimes {\mbox{\bf 1}}) 
\lambda{}^r (F_1\otimes ... \otimes F_r) 
\eeq
for all $F_1, ..., F_r$ in $\Lm$. 
Here $m$ is the multiplication map in 
$\Lm$,  $m(F_1\otimes F_2):=F_1 \we F_2$,  
and $\lambda{}^r$ is a linear map 
$\bigotimes{}^r \Lm \rightarrow \Lm \otimes \Lm$ 
given in terms of the map 
$\lambda: \Lm \rightarrow \Lm \otimes \Lm $: 
$\lambda(F) := F\otimes {\mbox{\bf 1}} - {\mbox{\bf 1}} \otimes F$ 
as follows: 
$\lambda{}^r (F_1\otimes ... \otimes F_r) :=
\lambda(F_1) \we ... \we \lambda(F_r) $.  
The graded differential operator $D$ is said to be of $r$-th 
order iff $\Phi{}^{r+1}_D = 0$ identically. For example, 
the identity (3.10) can compactly be written as 
$\Phi{}^{3}_{''}(a,b,c) = 0$.  

Now,  
the higher-order Leibniz rule for the (left) bracket with a $p$-form
$\ff{p}$ can be written as follows:
\beq
{\mbox{\Large $\Phi$}}^{n-p+1}_{\mbox{\small $\pbr{\ff{p}}{\,.\,} $} }
(F_1, ... , F_{n-p+1})=0.
\eeq
The simplest non-trivial
generalization corresponds to $p=(n-2)$. In this case the following 
(left) second-order graded Leibniz rule is obtained (cf. (3.10)) 
\beqa
\pbr{\ff{n-2}}{\ff{q}\we\ff{r}\we\ff{s}}
&=&\pbr{\ff{n-2}}{\ff{q}\we\ff{r}}\we\ff{s} 
+(-1)^{q(r+s)}\pbr{\ff{n-2}}{\ff{r}\we\ff{s}}\we\ff{q} 
\nn \\
&+& (-1)^{s(q+r)}\pbr{\ff{n-2}}{\ff{s}\we\ff{q}}\we\ff{r}
- \pbr{\ff{n-2}}{\ff{q}}\we\ff{r}\we\ff{s}
\\
&&\hspace*{-4.5em}-(-1)^{q(r+s)}\pbr{\ff{n-2}}{\ff{r}}\we\ff{s}\we\ff{q}
- (-1)^{s(q+r)}\pbr{\ff{n-2}}{\ff{s}}\we\ff{q}\we\ff{r} .
\nn
\eeqa
Thus, in addition to the structure of a noncommutative Gerstenhaber 
algebra we have here a higher-order generalization of the left graded 
Leibniz rule. 
This is another feature of field theory 
in 
the 
polymomentum formulation: 
the Poisson bracket 
can act as a higher order (algebraic) differential operator 
on the algebra of dynamical variables 
(here, the exterior algebra of forms), 
not only as a first order differentiation like  in mechanics.  
The structure given by the replacement 
of the graded derivation property of the bracket 
by the higher-order (left) graded Leibniz rule, eq. (3.12),   
can be naturally referred to as a 
{\em higher-order  (left) Gerstenhaber algebra. }

\section{The Nambu-type bracket}

\newcommand{\jac}[2]{\frac{\der(#1)}{\der(#2)}}
\newcommand{\jacd}[2]{\frac{d(#1)}{d(#2)}}

In this section we show that 
the polysymplectic form  
can also be used for defining 
the Nambu-type  analogue of the Poisson bracket.   
Here we only outline the idea 
in the particular case 
of  field theory in two dimensions,    
so that the  
polysymplectic form $\Omega$ is a three-form now. 
We can  choose 
either the form (1.2) corresponding to the DW canonical theory,  
or another appropriate non-degenerate closed three-form 
(see e.g. eq. (4.9) below) which 
 will  thus correspond to a certain  canonical 
theory from the Lepagean class of theories (cf.  e.g. \cite{lepage}).  

The polysymplectic form maps a function $F=F(y,p,x)$ 
of the polymomentum 
phase space variables to a bivector field $X_F$
\beq
X_F\inn\Omega= d F .
\eeq
Then the bracket of {\em three} functions can be defined 
as 
\beq
\{ F,G,K\}:=X_F\inn (dG\we dK) . 
\eeq
This bracket is antisymmetric in all three arguments 
and satisfies 
the Leibniz rule
\beq
\{ F,G,K\cdot L\}= \{ F,G,K\}\cdot L + \{ F,G,L\}\cdot K . 
\eeq
	Let us consider the analogue of the Jacobi identity for the 
	bracket above. Note first that the Nambu bracket (4.2) can be 
	related to the binary bracket  of a function with a one-form 
(cf. Sect. 2)  
\beq
\{F,G,K\}=X_F\inn (dG\we dK) = X_F \inn d(GdK)=\pbr{F}{GdK} . 
\eeq
Then the  Jacobi identity for the binary bracket, 
which is easy to  prove, 
can be used for deriving the  analogue of the 
Jacobi identity for the Nambu bracket:   
\beqa
\{\{F,G,H\}K,L\}&=&\pbr{\pbr{F}{GdH}}{KdL}=
\mbox{(Jacobi identity for $\pbr{}{}$)} \nn \\ 
&=& \pbr{F}{\pbr{GdH}{KdL}} - \pbr{GdH}{\pbr{F}{K dL}} , 
\eeqa
where the bracket of two one-forms $GdH$ and $KdL$ 
is  given by (cf. Sect. 3) 
$$
\pbr{GdH}{KdL}:= - X_{GdH}\inn d (KdL)  = - L_{X_{GdH}} (KdL)  
$$ 
and $X_{GdH}\inn \Omega = d (GdH)$. 
Now, let us express the first term in (4.5) in terms of  
the brackets of functions with one-forms:
\beqa
\pbr{F}{\pbr{GdH}{KdL}}
&=& X_F\inn d(-X_{GdH}\inn (dK\we dL))   
\nn \\
&=& X_F\inn d \left ( \pbr{GdH}{K} dL - dK \pbr{GdH}{L} \right ) \nn \\
&=&\pbr{F}{\pbr{GdH}{K}dL} - \pbr{F}{\pbr{GdH}{L}dK} .
\nn 
\eeqa 
Rewriting the result in terms of the Nambu bracket according to (4.4) 
and using the antisymmetry properties of the latter 
we obtain 
\beq
\{\{G,H,F,\}K,L\}+\{F,\{G,H,K\},L\}+\{F,K,\{G,H,L\}\}=\{G,H,\{F,K,L\}\} .
\eeq
The identity above is known as the ``fundamental identity'' for the Nambu 
bracket \cite{takh} 
which  plays the role of an  analogue of the 
Jacobi identity for the 
latter. It shows that 
the Nambu bracket 
introduced above endows the space of functions on the polymomentum phase 
space with the structure of what could be called 
 a  
{\em Nambu-Poisson algebra}.

In order to write the equations of motion of dynamical variables 
in terms of the Nambu 
bracket let us first observe that given an appropriate  
polymomentum Hamiltonian function $H$ 
we can associate with it the 
``Hamiltonian flow'' which is a distribution of two-planes given 
by the equation 
\beq
\jac{z^M,z^N}{x^1,x^2}=X_H^{MN}(y,p,x) .
\eeq
In the particular case when $H$ is chosen to be the 
DW Hamiltonian function and $\Omega$ is given by (1.2) 
this equation reproduces the De Donder--Weyl Hamiltonian form 
of field equations, eq. (1.1). 
{}From (4.7) it follows that the bracket with $H$  
determines   
{\em joint} equations of motion of two 
dynamical variables: 
\beq
\jacd{F,G}{x^1,x^2}=\{ H,F,G\}+\jac{F,G}{x^1,x^2}, 
\eeq
where the ``total jacobian'' is introduced: 
$$\jacd{F,G}{x^1,x^2}:=\half\jac{F,G}{z^M,z^N}\jac{z^M,z^N}{x^1,x^2}.$$  

\newcommand{\ngoto}{
As an example let us consider the Nambu-Goto string. 
Here $y^1,...,y^D$ are to be the target space coordinates and 
$x^1,x^2$ are to be the string world-sheet parameters. 
For the seek of simplicity let us put the string tension $T=1$.  
>From the  Lagrangian density 
 
we determine the polymomenta 

and the DW Hamiltonian function 
which can be written in terms of 
the polymomenta (cf. \cite{}) 
Using the polysymplectic form () and the map () we 
can calculate the components of the bivector 
$X_H$: the non-vanishing components are 
$X^{ai}= ... \epsilon^{ij}\der H /\der p_a^j$. 
The string equations of motion can be written now as follows 
\beqa
\jacd{y^a,x^i}{x^1,x^2}=\{H,Y^a,x^i\} = ... \epsilon^{ij} \der H /\der p_a^j \\
\jacd{p^k_a,x^i}{x^1,x^2}=\{H,p^k_a,x^i\} = 0 
\eeqa
It is easy to check that there are equivalent to the Nambu-Goto 
equations of motion. 
}

As we have mentioned above 
Nambu brackets can be defined within the  polymomentum 
canonical theories different from  the De Donder--Weyl theory. 
All these theories are known to 
follow from the general Lepage framework in the calculus of variations 
of multiple integral problems (see e.g. \cite{lepage}). 
The difference between 
them results after all in a different polysymplectic form and a 
different definition of the polymomenta and the analogue of 
the canonical Hamilton's function $H$.    

For instance, in the case of the Schild (or Nambu-Goto) 
string a canonical theory can be developed 
which is based on the 
following analogue of the polysymplectic form 
(cf. \cite{beig}, see also \cite{nambu80})   
\beq 
\Omega_C:=d p_{ab} \we dy^a \we dy^b .
\eeq 
Here, new polymomenta $p_{ab}$  are given by 
$p_{ab}:= \der L / \der v^{ab}$,  
where the jacobian  
  $v^{ab}:=\der(y^a,y^b)/\der(x^1,x^2)$ 
  naturally plays the role of a 
two-dimensional analogue of  ``generalized velocities''. 
The analogue of the Hamiltonian function, $H_C$, 
which is now a function of  new 
polymomentum phase space variables $(y^a, p_{ab}, x^i)$,   
is naturally  defined  as follows: \mbox{$H_C=p_{ab} v^{ab} - L$}.  
For the Schild string 
with $L = \frac{1}{2}v_{ab}v^{ab} $   
(for the seek of simplicity we put the string tension $T=1$) 
we 
obtain $H_C= \frac{1}{2} p_{ab} p^{ab}$. 
 The form $\Omega_C$  maps functions 
to bivectors and gives rise to the Nambu bracket of three functions 
as in (4.2). The equations of motion 
written in terms of the Nambu bracket assume the form 
\beq 
\jacd{y^a,y^b}{x^1,x^2}=\{H_C, y^a,y^b \} = p^{ab},    
\quad  \jacd{p_{ab},y^b}{x^1,x^2}=\{H_C,p_{ab},y^b\}=0.  
\eeq 
It is easy to show that  these equations are equivalent to the 
equations of motion  of the Schild string 
(see e.g. \cite{nambu80}).

In general, in $n$-dimensional field theory 
an $(n+1)$-ary Nambu bracket can be defined. 
In mechanics ($n=1$) this bracket 
reduces to the usual binary Poisson bracket. 
In this sense 
the Nambu bracket  is a generalization of 
the Poisson bracket to field theory 
(within the polymomentum formulations). 
	In this case, the Nambu bracket 
	with a properly defined polymomentum 
	analogue  of Hamilton's canonical function 
	determines a {\em joint} evolution of $n$ dynamical variables, 
contrary to the 
common point of view in  
generalized Nambu mechanics  
in which  the evolution of a dynamical 
variable 
is supposed to be 
given by the bracket with several ``Hamiltonians'' 
(see e.g. \cite{nambu,takh}).  

\section{Conclusion}

When trying to extend the notion of the Poisson bracket to the 
polymomentum formulations 
in field theory, we immediately face the 
problem related to the asymmetry between the number of 
field variables and the polymomenta. 
One way out is to 
define a bracket on differential forms instead of functions. 
Another way is to consider an $(n+1)$-ary bracket of 
the Nambu type instead of 
a binary bracket. 
In these two approaches the role of dynamical variables 
in field theory 
is played by forms and functions, respectively 
(not functionals as in the 
standard Hamiltonian formalism).  
Here we have considered the algebraic structures which appear 
within both of these approaches. 
We have shown that 
on a certain subspace of forms, called Hamiltonian, 
the bracket of forms gives rise 
to a Gerstenhaber algebra, whereas   
on arbitrary horizontal forms 
its generalizations -- noncommutative (in the sense of Loday) 
and higher-order (in the sense of the higher order Leibniz property 
of the bracket) -- appear. 
On the other hand, 
the $(n+1)$-ary bracket of functions fulfills all the properties of the 
Nambu  bracket including the famous fundamental identity [10],  
thus leading to a Nambu-Poisson algebra.   
All the abovementioned structures reduce to the familiar Poisson algebra 
when $n=1$, that is in mechanics. In this sense they all 
are generalizations of a Poisson algebra to field theory. 
An interesting problem for future research would be the   
development of a quantization scheme in field theory 
based on the structures discussed above.  

\medskip

{\bf Acknowledgements. }   
I would like to thank B. De Witt and G. Sardanashvily for useful discussions 
at the Bialowie\.za'96 Workshop.  
I am also grateful to C. Roger for drawing my  
attention to  Loday's paper \cite{loday}.


\end{document}